\documentclass[aps,prb,superscriptaddress,twocolumn,amsmath,amssymb,amsfonts,floatfix,showpacs]{revtex4-1}
\usepackage{graphicx,color,bm}
\usepackage{graphicx}
\usepackage{amsmath}
\usepackage{CJK}
\usepackage[colorlinks=true,citecolor=blue,urlcolor=blue]{hyperref}
\usepackage{natbib}
\usepackage{times}
\begin{document}
\begin{CJK}{UTF8}{bsmi}
\title{Half-Magnetization Plateau of a Dipolar Spin Ice in a [100] Field}
\author{Sheng-Ching Lin(林昇慶)} 
\affiliation{Department of Physics and Center of Theoretical Sciences, National Taiwan University, Taipei 10607, Taiwan}

\author{Ying-Jer Kao(高英哲)}
\email{yjkao@phys.ntu.edu.tw}
\affiliation{Department of Physics and Center of Theoretical Sciences, National Taiwan University, Taipei 10607, Taiwan}
\affiliation{Center for Advanced Study in Theoretical Science, National Taiwan University, Taipei 10607, Taiwan}
\date{\today}

\begin{abstract}
We report here numerical results of the low-temperature behavior of a dipolar spin ice in a  magnetic field along the [100] direction.  Tuning the magnetic field, the system exhibit a half-magnetization plateau at low temperature.  This half-polarized phase should correspond to a quantum solid phase in an effective 2D quantum boson model, and the transition from the  Coulomb phase with a power-law correlation to this state can be regarded as a superfluid to a quantum solid transition. We discuss possible experimental signatures of this half-polarized state. \end{abstract}

\maketitle
\end{CJK}

Topological phases of matter is one of the most fascinating phenomena  in condensed matter systems.  Fractional quantum Hall states,\cite{Wen:1995kx} spin liquids,\cite{Balents:2010ce} and recently proposed topological insulators\cite{Qi:2011fk} are some well-known  examples.  Spin ice, a  geometrically frustrated magnet with Ising spins on a pyrochlore lattice of corner-sharing tetrahedra,\cite{Bramwell:2001fk}, at low temperatures exhibits a ``Coulomb'' phase with dipolar correlation and fractionalized monopole excitations.\cite{Ryzhkin:2005fk,Castelnovo:2007eh,Henley:2010fk} This phase emerges due to a local constraint,  the ``ice rule'',  with  two spins pointing in and two out of each tetrahedron. The ice rule  can be regarded as a conservation law of an emergent gauge field,\cite{Ryzhkin:2005fk,Castelnovo:2007eh,Henley:2010fk} and it gives rise to the dipolar correlation among spins. This local constraint leads to an  extensive ground state degeneracy, and at low temperature, the spin ice has an extensive residual  entropy, $S_0 \approx (k_B/2)\ln(3/2)$ per spin, first estimated by Pauling for the water ice.\cite{Pauling:1935dq}

Starting from the nearest-neighbor spin ice  (NNSI) and adding perturbation such as an external magnetic field or  further-neighbor dipole-dipole interaction, the system can undergo a phase transition from the Coulomb phase to an ordered state at sufficiently low temperature. This confinement  transition is highly unconventional, and can not be described by the  Landau-Ginzburg-Wilson paradigm. The transition to a non-magnetic state  can be regarded as a Higgs transition involving condensation of an emergent matter field coupled to the U(1) gauge field.\cite{Powell:2011uq} Applying an external magnetic field along the [100] direction, the transition from an ordered  $\mathbf{q}=0$ fully polarized (FP) state to a Coulomb phase corresponds to  a three-dimensional Kasteleyn transition, which is  first-order like on the one side and continuous on the other.\cite{Jaubert:2008dq} This transition corresponds to a proliferation of  string excitations along the [100] direction, which  corresponds to the condensation of bosons in an effective two-dimensional hardcore boson model through a classical to quantum mapping,\cite{Powell:2008fk,*Powell:2012uq,*Powell:2013fk}

On the other hand, in  real spin ice materials, such as Dy$_2$Ti$_2$O$_7$ (DTO) and Ho$_2$Ti$_2$O$_7$ (HTO), the rare earth ions have large magnetic moments ($\sim 10\mu_B$),\cite{Gardner:2010xv} and the long-range dipole-dipole interaction can not be ignored. The low-temperature states of the dipolar spin ice (DSI) also satisfy the ice rule, and it is shown that the NNSI and the DSI are projectively equivalent apart from small interactions decaying with the separation, $r$,  faster than $1/r^3$.\cite{Isakov:2005qf} At very low temperature, this residual interaction selects a $\mathbf{q}=(001)$ long-range order, and the Pauling's residual entropy  is released through a first-order phase transition.\cite{Melko:2001fk,*Melko:2004uq} This Melko-den Hertog-Gingras (MDG) phase is non-magnetic and has so far escaped experimental observation, although recent specific heat measurement in a  thermally equilibrated  DTO sample  shows  possible signatures of such an ordered phase.\cite{Pomaranski:2013uq}

\begin{figure}[bp]
\centerline{
\includegraphics[width=3in]{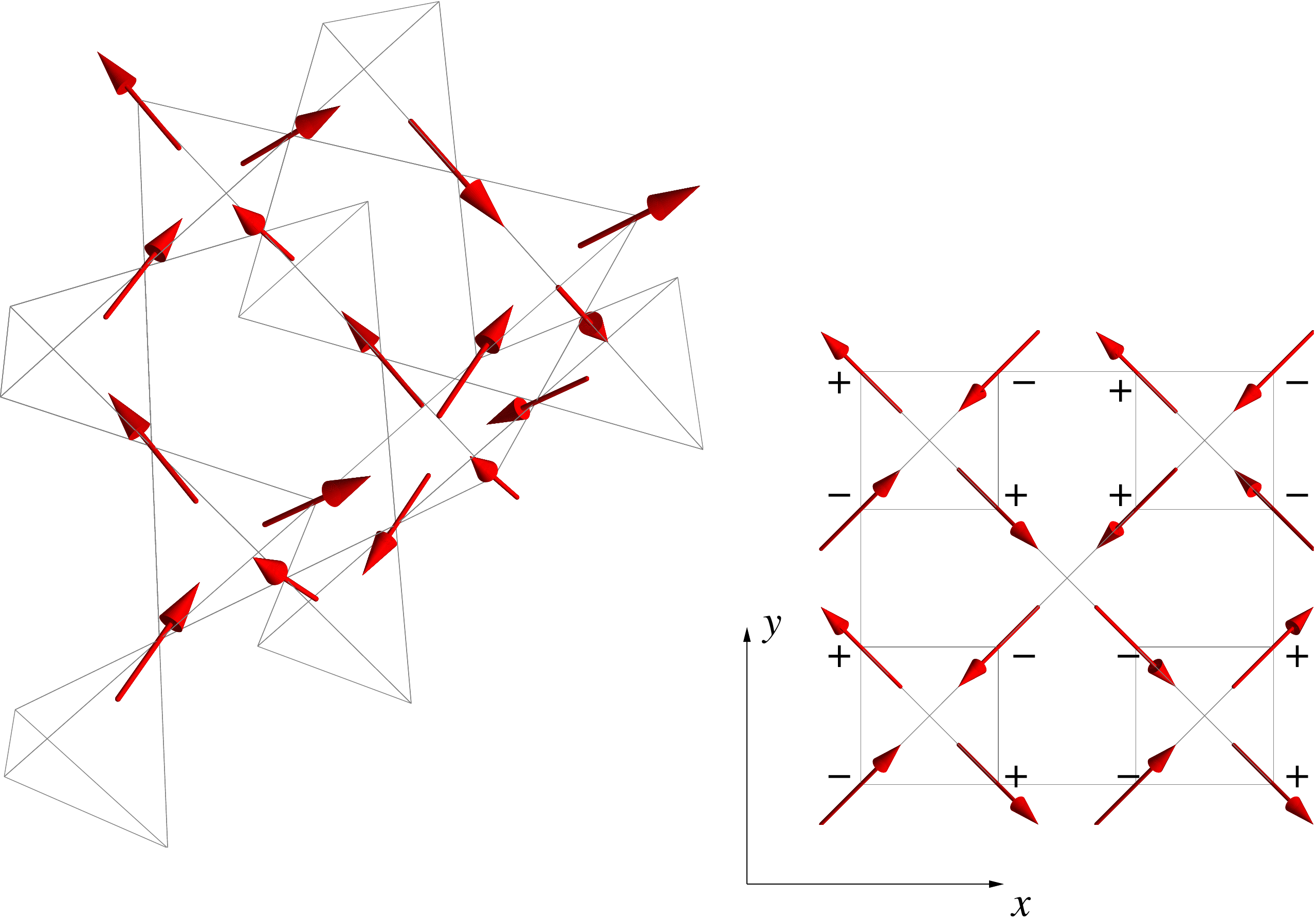}}
\caption{(Color online). (Left) Pyrochlore lattice showing the HP state. (Right) The HP  state projected along the $z$-axis.  The component of each spin parallel to the $z$ axis is indicated by $+$ and $-$ sign.  }
\label{fig:spinin}
\end{figure}

The natural question arises: What kind of  low-temperature ordered phase will emerge if one applies a [100] magnetic field in a DSI? How to characterize the phase transition from the disordered Coulomb phase to the ordered state? In this {Letter}, using large-scale Monte Carlo (MC) simulations of the DSI in a [100] magnetic field,  we find a new ordered state with half of the saturated magnetization stabilized by the dipolar interaction. Starting from the MDG phase with zero magnetization at low field, the DSI  transitions to the half-polarized (HP) state upon the increase of the magnetic field, before the system finally reaches an FP state at high field. This should be contrasted with the half-magnetization plateau found in the Heisenberg model, which is stabilized  by either the lattice  distortion,\cite{Penc:2004uq} or the quantum effects.\cite{Bergman:2006fk} Through a classical to quantum mapping,\cite{Powell:2008fk,Powell:2013fk} the HP state should correspond to a quantum solid (QS) in a two-dimensional (2D)  quantum boson model, and the Coulomb  to the HP state transition would correspond to a superfluild (SF) to a QS transition.\cite{Hebert:2001fk} Signatures of this HP state may have been observed in the magnetization and neutron scattering measurements of DTO and HTO.\cite{Fennell:2005fk}

\textit{Model and Method.}-- The Hamiltonian of a dipolar spin ice  in an applied [100] magnetic field is given by 
\begin{align}
H&=\frac{J}{3}\sum_{\langle(i,a),(j,b)\rangle}\sigma^{a}_{i}\sigma^{b}_{j}\nonumber \\
&+Dr_{nn}^3\sum_{i<j,a,b}\left(\frac{{\mathbf{n}^a}\cdot\mathbf{n}^b}{|\mathbf{R}_{ij}^{ab}|^3} -\frac{3(\mathbf{n}^a\cdot\mathbf{R}_{ij}^{ab})(\mathbf{n}^b\cdot\mathbf{R}_{ij}^{ab})}{|\mathbf{R}_{ij}^{ab}|^5} \right)\sigma^{a}_{i}\sigma^{b}_{j}\nonumber \\
&- \frac{B}{\sqrt{3}}\sum_{i,a}\sigma_i^{a'}, 
\label{eq:H}
\end{align}
where $i,j$ are the tetrahedron indices, $a,b$  corresponds to the sub-lattices inside each tetrahedron and $\mathbf{n}^a$ is the local $\langle 111 \rangle$ axis. The Ising variable $\sigma=\pm 1$ denotes spin pointing into or out of the tetrahedron and  $\sigma'$ denotes the sign of the $z$-component of the spin.  $J$ is the NN exchange interaction, $D=(\mu_0/4\pi)\mu^2/r_{nn}^3$ is the  strength of the dipolar interaction, and $r_{nn}$ is the pyrochlore NN distance.    The long-ranger dipolar interaction is taken into account  using the Ewald method to avoid any truncation effects. In the following, we will use the parameters for DTO with  $\mu\approx 10\mu_B$, $r_{nn}=3.54$\AA, $J = -1.24$K, and $D=1.41$K.\cite{Hertog:2000ly} 

We perform classical Monte Carlo simulations on this model in a conventional cubic unit cell of edge size $L$ with a  total number of spins  $N=16L^3$. We remain in the spin-ice manifold with two-in-two-out spin configurations on each tetrahedron and update  the configuration using  both the  loop and   worm algorithms. \cite{Barkema:1998fk,Melko:2001fk,Melko:2004uq, Jaubert:2008dq,Jaubert:2009nx} The spin configurations remain in the spin-ice manifold after these updates. In order to incorporate the long-range dipolar interaction in the  worm updates, we first generate a loop using the worm algorithm for the NNSI,\cite{Jaubert:2008dq,Jaubert:2009nx} and then compute the energy change $\Delta E_{\rm dip}$ from the dipolar interaction. We then flip the spins on the loop with the Metropolis probability, 
$P=\min (1, \exp(-\beta\Delta E_{\rm dip}))$. The conventional loop update is necessary as the acceptance rate of the worm updates drops significantly at low temperatures.\cite{SM} 
In addition to these updates, we perform parallel tempering in either the temperature  or  the magnetic field domain to  avoid freezing in the algorithm.\cite{Earl:2005kx} One MC step  in our simulation consists of loop moves of  each type, and we perform parallel tempering swap after 100 MC steps. 
In a typical simulation,  at least $6\times10^6$ MC steps is carried out for equilibration  at each temperature(field), and another $6\times 10^6$ MC steps for data production. The temperature(field)  intervals in the tempering scheme are selected to achieve constant acceptance rates for swaps.  Due to the long-range nature of the dipolar interaction, the computation scales as $O(N^2)$, and our largest simulation is thus limited to $L= 5$ with $N=2000$ spins. Although this size is small compared to a largest  NNSI simulation with $L=100$,\cite{Jaubert:2008dq} $L=5$ is the largest-size simulation for a DSI.\cite{Melko:2004uq,Lin:fk} In our simulations, we do not observe strong size dependence for $L=4$ and 5; however, we can not rule out the possibility that there might exist lower energy configurations which can not fit into our simulation cell.    

\begin{figure}[tbp]

\centerline{\includegraphics[width=0.5\textwidth]{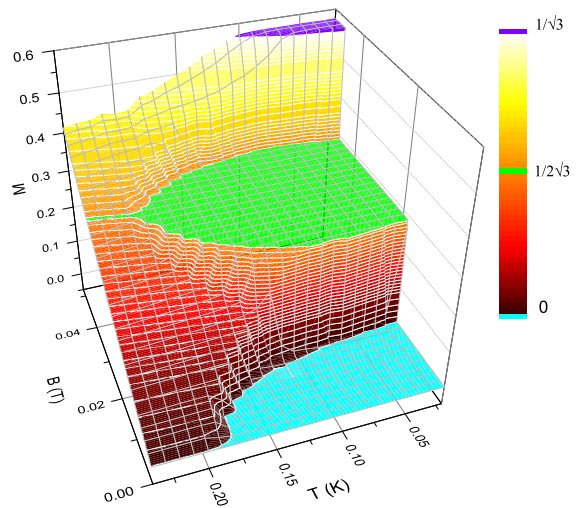}}
\caption{(Color online). Magnetization (in units of 10$\mu_B$) versus temperature and  magnetic field. Three magnetization plateaus exist at low temperature, which corresponds to three ordered states: the MDG state (cyan), the HP state (light green) and the FP state (purple).  Notice the temperature axis is inverted for  clarity.}
\label{fig:ML5}
\end{figure}

\begin{figure}[tbp]
\begin{center}
\includegraphics[width=\columnwidth,clip]{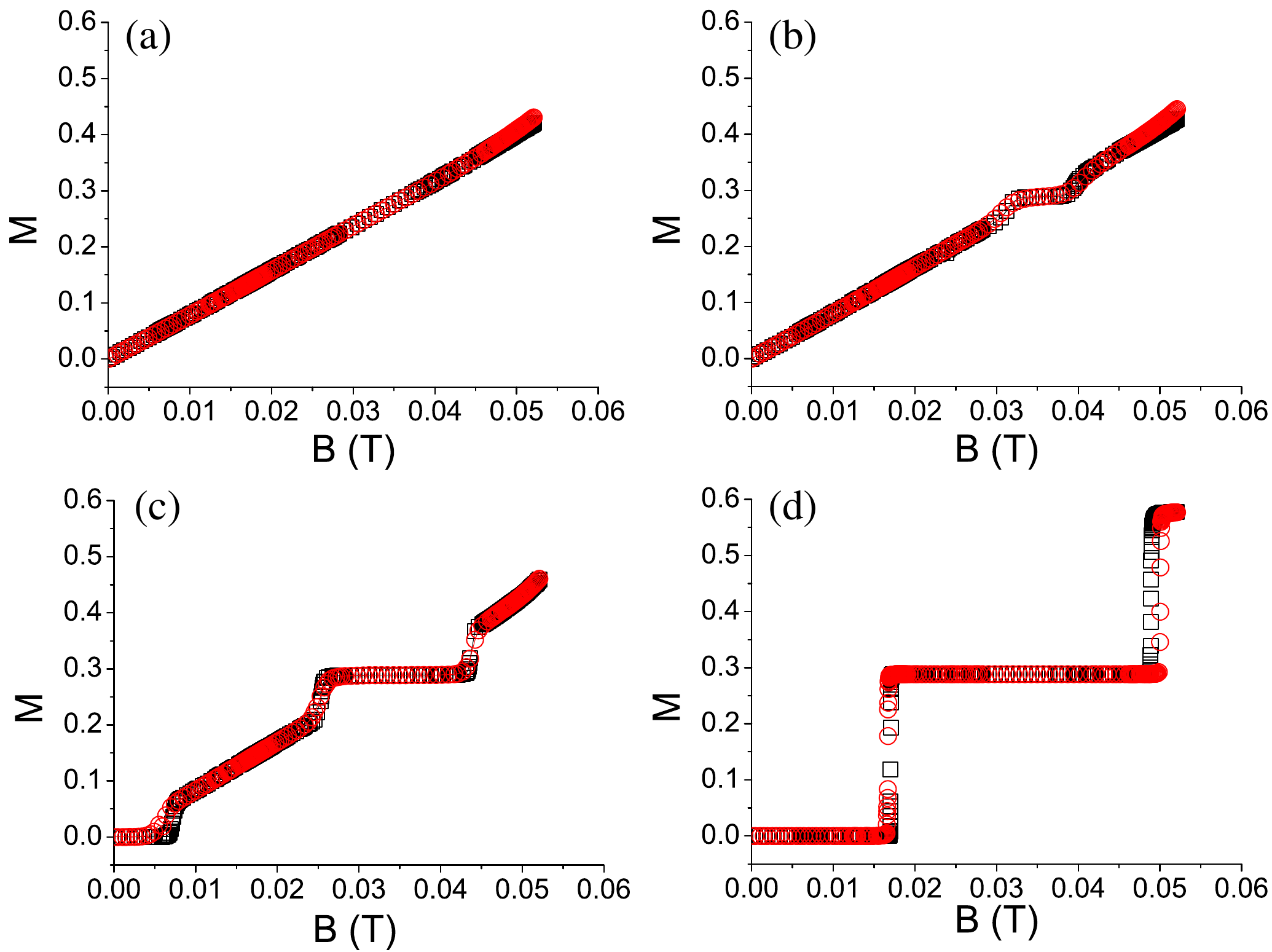}
\end{center}
\caption{(Color online). Magnetization (in units of 10$\mu_B$) curves at $T=$(a) $0.22$K, (b) $0.20$K, (c) $0.18$K, and (d) $0.07$K for $L$=4 (black square) and 5  (red circle).  }
\label{fig:fig3}
\end{figure}

\textit{Results and Discussion.}-- Figure~\ref{fig:ML5} shows the magnetization as a function of magnetic field and temperature. There exist three magnetization plateaus at  low temperature: the MDG state with zero magnetization, the HP state with half of the saturated magnetization, and the FP state. Upon raising the temperature, the half-magnetization plateau becomes less stable and eventually disappears at high temperature. Figure~\ref{fig:fig3} shows the details of the magnetization curves at different temperatures. At $T=0.22$K, the magnetization  curve shows a collective paramagnetic behavior without any signature of a plateau, and  the system is in the disordered Coulomb phase.  At $0.20$K, a small half-magnetization plateau  appears near $B=0.035$T. The crossover becomes more step-like as the temperature is lowered, and the zero magnetization plateau of the MDG state emerges at small field. At $0.07$K,  sharp plateau transitions  are observed.  These behavior can be understood by considering the field dependence of  the energy for the  three ordered states: \cite{SM} the MDG state is non-magnetic, the HP state has half of the saturated moment, and the FP state has the saturated moment, and they have different field dependence; therefore, there exist two level crossings as one tunes the magnetic field. These level crossings give rise to the plateau transitions at low temperatures. The high temperature crossovers can be also understood  from  the  free energy of the system, and we refer the readers to the Supplemental Materials for details.\cite{SM}

Closer examination of the spin configurations at the half-magnetization plateau, we find the configuration can be regarded as a half-half mixture of the non-magnetic MDG and the saturated FP states (Fig.~\ref{fig:spinin}). It contributes to half of the saturated magnetization since two of the tetrahedra in the cubic unit cell have magnetic moments along the direction of applied field, similar to those from the saturated FP state. The magnetic moments from the other two tetrahedra cancel  as in the MDG state.  

\begin{figure}[tb]
\includegraphics[width=0.9\columnwidth,clip]{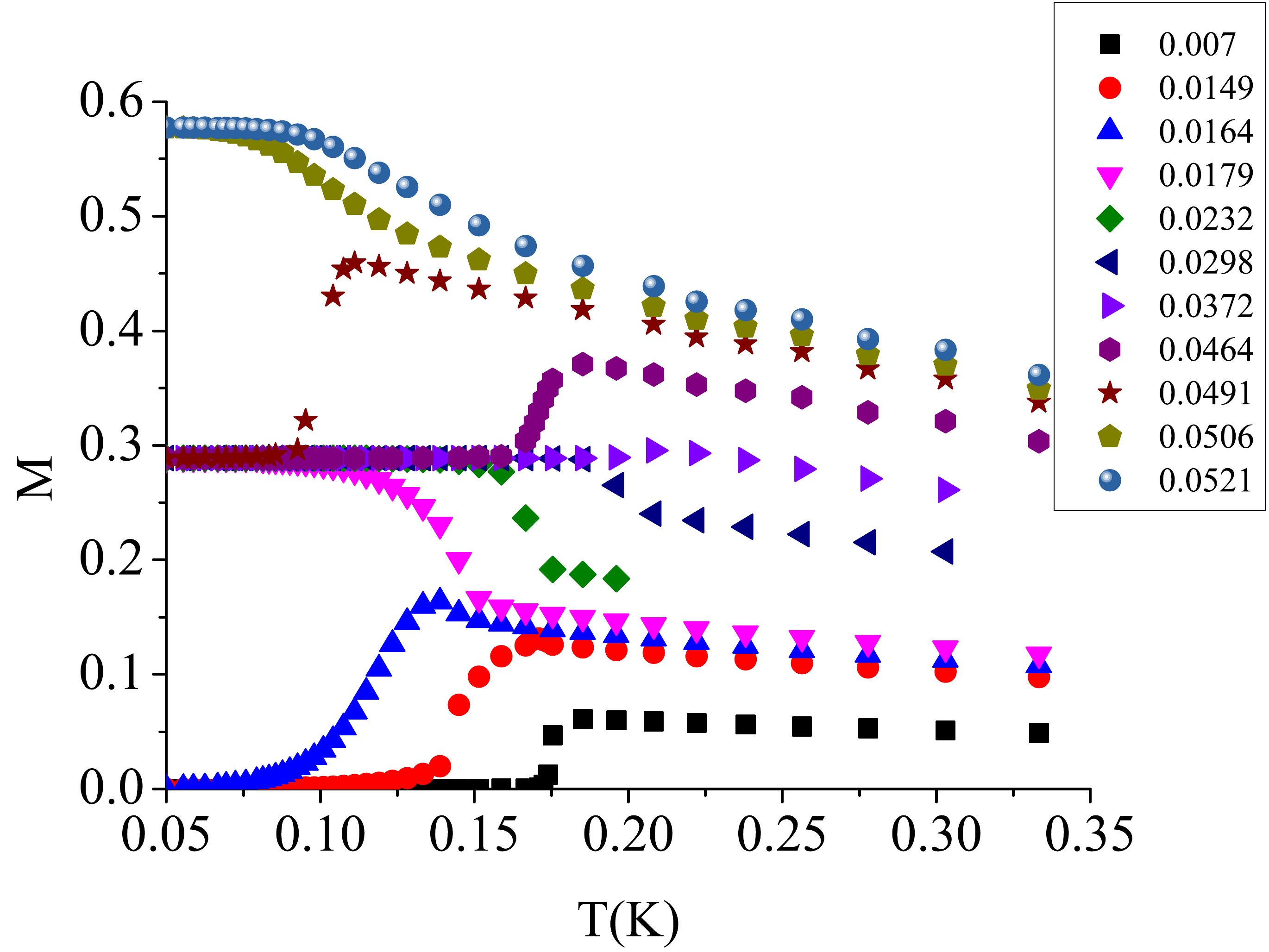}
\caption{ (Color online)
 Temperature dependence of the magnetization at different magnetic fields along the [100] direction. Magnetization is in units of 10$\mu_B$, and the magnetic fields are in Tesla. }
\label{fig:MT}
\end{figure}

Figure~\ref{fig:MT} shows the temperature dependence of the magnetization at different magnetic field strength. At low fields $B<0.017$T, the systems enters from the Coulomb phase to the MDG phase with zero magnetization at low temperatures. For intermediate fields, the system transitions into the HP state. At fields larger than 0.05T, a rounded Kasteleyn transition from the disordered Coulomb phase to a FP $\mathbf{q}=(000)$ state is observed.\cite{Jaubert:2008dq,Morris:2009qf} This rounding might be either due to the finite-size effect, or the high magnetic field. 
The low-temperature ground state is very sensitive to the delicate balance between the magnetic field and the dipolar interaction. Two magnetization curves with slightly different magnetic fields may have similar temperature dependence at high temperature; however, the  states at low temperature will settle to different magnetization plateaux (Fig.~\ref{fig:MT}) .

 The thermal phase transition from the Coulomb phase to the ordered states show sharp specific heat anamoly  due to the strongly first-order nature of the transition, and the large residual entropy is released at the transition.\cite{SM} In contrast, the transitions between the ordered states show relatively small specific heat anomaly as these are  ordered states with different magnetizations. Experimentally, it has been difficult to observe the MDG phase due to the increase of the spin relaxation time as temperature is lowered, and the system is hard to reach  thermal equilibrium. \cite{Matsuhira:2011fk,*Revell:2013uq,*Yaraskavitch:2012kx} On the other hand, the FP state can be easily reached at high temperature and  in large field. \cite{Fukazawa:2002fk} Here we propose another route to reach the MDG phase: First, the system is cooled down in  a large [100] field from the high temperature disordered Coulomb phase to the FP state. Then, the magnetic field is lowered at low temperature. The transition from the FP state to the MDG state through an intermediate HP state can be more easily reached experimentally since they go through transitions between ordered states.

\begin{figure}[tb]
\centerline{\includegraphics[width=\columnwidth]{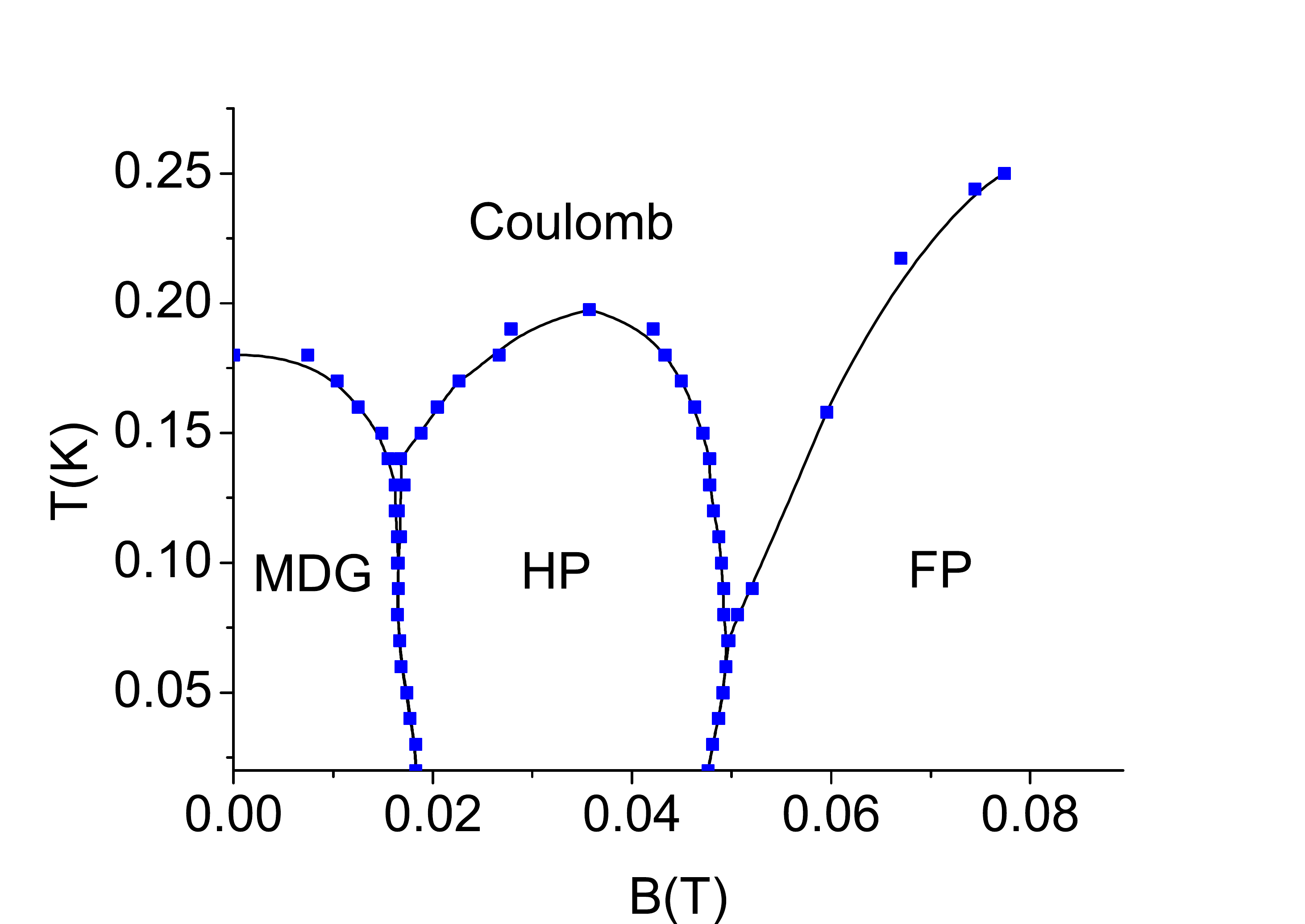}}
\caption{ The phase diagram of the dipolar spin ice in a [100] field. Blue squares are the  transition points estimated from the Binder ratio. The solid curves are drawn as a guide for the eye. See text for description of the phases. }
\label{fig:Dip_D}
\end{figure}

Figure~\ref{fig:Dip_D} shows the phase diagram of the DSI  for fields applied along the [100] direction, using the DTO parameters. The phase boundaries are estimated using the Binder ratio (blue squares).\cite{Binder:1981fk} At  high temperature, the system is in the disordered Coulomb phase with dipolar  correlations.\cite{Castelnovo:2007eh}  As the temperature is lowered, the combination of the magnetic field and the dipolar interaction  will select either the MDG or the HP  state.  It should be noted that there exists slight reentrant behavior near the phase boundaries between the MDG and the HP phases, and between the HP and the MDG phases. The thermal transitions from the Coulomb to  the MDG phase, and to the HP phase are first-order, and they merge at  the tip of the HP lobe as a (putative) critical point.  Through  the quantum mapping,\cite{Powell:2008fk} the MDG phase corresponds to a MI, the HP phase  corresponds to a half-filled checkerboard QS (density wave) in a 2D extended Bose-Hubbard model,\cite{Hebert:2001fk} and the Coulomb phases corresponds to a superfluid (SF) phase. The transition from  the Coulomb phase to an ordered state thus corresponds to a SF-MI or SF-QS transition in the quantum model. In this mapping, the [100] field $B$  is proportional to $\mu$, and the magnetic moment $M$ would correspond to the average boson density $\rho$. Interestingly, the magnetization (Fig.~\ref{fig:ML5}) and the phase diagram (Fig.~\ref{fig:Dip_D}) show great resemblance to their counterparts in a 2D extended hardcore Bose-Hubbard model on a  square lattice,\cite{Hebert:2001fk} 
\begin{align}
H&=-t\sum_{\langle ij\rangle} \left(b^\dagger_i b_j+ \mbox{h. c.}\right) +V\sum_{\langle ij\rangle}  n_i n_j -\mu \sum_i n_i,
\label{eq:bose}
\end{align} 
where $b_i$ and $b^\dagger_i$ are the annihilation and creation operators of hardcore bosons on site $i$, $n_i$ is the boson density, $\mu$ is the chemical potential, and $V$ is the nearest-neighbor  repulsive interaction. The HP phase  can be identified as  a half-filled checkerboard QS (density wave) in this model.
The SF-QS transition in the bosonic model is generally first-order as the two phases break different symmetries,\cite{Hebert:2001fk}  and similarly for the phase transition from the Coulomb to HP phase. Further analysis based on symmetries is required to construct an effective quantum model. \cite{Powell:2008fk,*Powell:2011uq,*Powell:2012uq,*Powell:2013fk}

What are the possible experimental signatures of the HP state? One should observe a clear half-magnetization plateau at low temperatures in  magnetization measurements.  The systems should also exhibit magnetic susceptibility anomalies near the plateau transitions.\cite{SM} Since the HP phase is a  mixture of the $\mathbf{q}=(000)$ and $(001)$ states, one should observe in neutron scattering two peaks at $(000)$ and (001), and the spectral weights for each peak  should be a fraction of that observed in  the MDG  and the FP phase, respectively. The  available magnetization data show no signs of the magnetization plateau at temperature as low as 0.6K and the system remains paramagnetic up to saturation.\cite{Fukazawa:2002fk,Morris:2009qf} On the other hand,  the integrated intensity at $(000)$ obtained from neutron scattering  on DTO  at $T$=0.05K shows steps with hysteresis at  $B \sim 0.2$T~ \cite{Fennell:2005fk}. More interestingly, the ordered moment in HTO  at $T$=0.05K shows a plateau of 6$\mu_B$  near $B\sim 0.4$T. Also, the diffuse scattering at $(001)$ persists in finite field and shows a steplike decrease in commensuration with the steplike increase in the $(000)$ scattering between $B\sim 0.2$ to 0.5T (Fig.~8 in Ref.~ \onlinecite{Fennell:2005fk} ).  These features have been  previously attributed to the  lack of relaxation of the magnetic moments at low temperatures, and the system is trapped in a metastable state.  On the other hand, these  features resemble the signatures of  a HP state, although the field range observed experimentally  is  one order of magnitude larger than our prediction, and the signatures disappear upon cycling through the field. \cite{Fennell:2005fk} Further experiments are necessary to determine whether these indicate the existence of  the HP phase.

\textit{Conclusion.}-- Using  large-scale Monte Carlo simulations, we construct the full phase diagram of a DSI in a magnetic field along the [100] direction. We find  a new intermediate HP phase  with half of the saturated magnetization, which should correspond to a QS phase in a 2D quantum boson model. The transition from the Coulomb phase to this phase thus corresponds  to a SF-QS transition in the bosonic model.  This points to  a new direction which can be further explored through the classical to quantum mapping. More theoretical analysis  based on symmetry considerations has to be done to determine  the general form for the quantum Hamiltonian, and the corresponding continuum quantum field theory  that  describes the characteristics  of the phase transition, in particular near the putative critical point at the tip of the HP phase. In other words, studying the purely classical DSI in a magnetic field may provide useful information regarding the quantum phase transitions  in a 2D quantum boson model,\cite{Hebert:2001fk} and  exotic phases such as  the supersolid phase in the lattice boson model may also be realized in the DSI.\cite{*[{See, for example, }] [{ and references therein.} ] Chen:2008kq} On the experimental side,  we believe the data presented in Ref.~\onlinecite{Fennell:2005fk} indicates possible signatures of this phase in both DTO and HTO, and our results will stimulate further experimental efforts to resolve this issue.  It will also be interesting to see what are the quantum effects on the magnetic plateau in the quantum spin ice.\cite{Ross:2011uq,*Chang:2012kx,*Wan:2012vn}



\begin{acknowledgments}
We thank  L. D. C. Jaubert for  a discussion on the improved worm algorithm, and S. Powell  for very useful comments. We also thank  J. T. Chalker and R. Moessner for helpful discussions. Y. J. K. thanks the hospitality of  the Aspen Center for Physics, where part of this work was done. This work is partially supported by the NSF under Grant No.~PHY-1066293, the NSC in Taiwan under Grants No.~100-2112-M-002-013-MY3, 100-2923-M-004-002 -MY3 and  102-2112-M-002-003-MY3, and by NTU Grant No.~101R891004. Travel support from NCTS in Taiwan is also acknowledged. 

\end{acknowledgments}


\appendix


\section{Comparison of loop and worm updates}

\begin{figure}[tbp]
\includegraphics[width=0.95\columnwidth,clip]{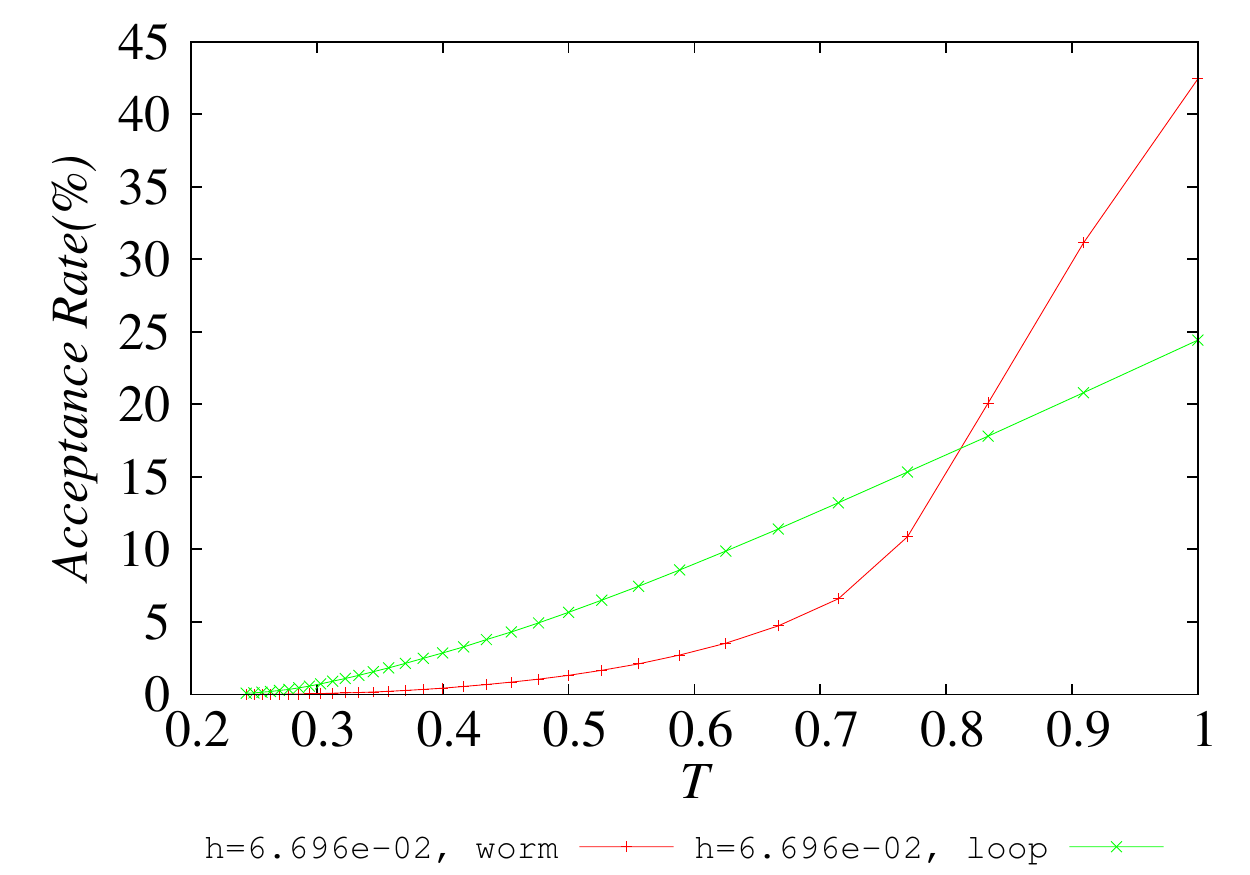}
\caption{(Color online) Comparison of the  conventional loop and directed loop updates at $B=6.696\times 10^{-2}$T. Worm updates becomes less efficient below $T=0.81$K. }
\label{fig:acc}
\end{figure}

Two types of cluster updates were used in our simulation: the loop \cite{Melko:2004uq} and worm algorithms \cite{Jaubert:2008dq,Jaubert:2009nx}. In the loop algorithm, a loop is randomly generated and is flipped with a Metropolis move.  It depends strongly on the energy difference involved between the initial and proposed states of the the loop move and can possibly be rejected \cite{Melko:2004uq}. For simulations in a field, this algorithm becomes less efficient as the rejection rate can be very high. On the other hand, in the worm algorithm, the detailed balance condition is imposed  along the construction of the loop, and it allows  to flip a long loop of spins in a field without rejection for the NN spin ice model \cite{Jaubert:2008dq,Jaubert:2009nx}. In our implementation of the worm algorithm, the loop is generated using the transition probabilities specified in the nearest-neighbor spin ice (NNSI) model, and the loop is flipped with the Metropolis probability $P=\min (1, \exp(-\beta\Delta E_{\rm dip}))$, where $\Delta E_{\rm dip}$ is the energy change due to the dipolar interaction. Figure~\ref{fig:acc} shows the acceptance rates of the loop and the worm updates at a field of $B=6.696\times 10^{-2}$T. 
The worm update is more efficient  at high temperatures since the loop can be  generated  without any backtracks  \cite{Jaubert:2008dq,Jaubert:2009nx}. 
However, the worm update become inefficient when temperature is lower than $0.81$K. This temperature  roughly coincides with the Kasteleyn transition temperature $T_k$ = $\frac{2h}{\sqrt3\ln2}$ for the NNSI, below which the backtracking must be included to ensure a physically meaningful solution for the transition probability, and the loop may terminate prematurely. On the other hand, the  loop in the loop update is generated by random walks  through tetrahedra \cite{Melko:2004uq}, and there still exists  finite probability to update the system.
\begin{figure}[tbp]
\centerline{\includegraphics[width=0.95\columnwidth,clip]{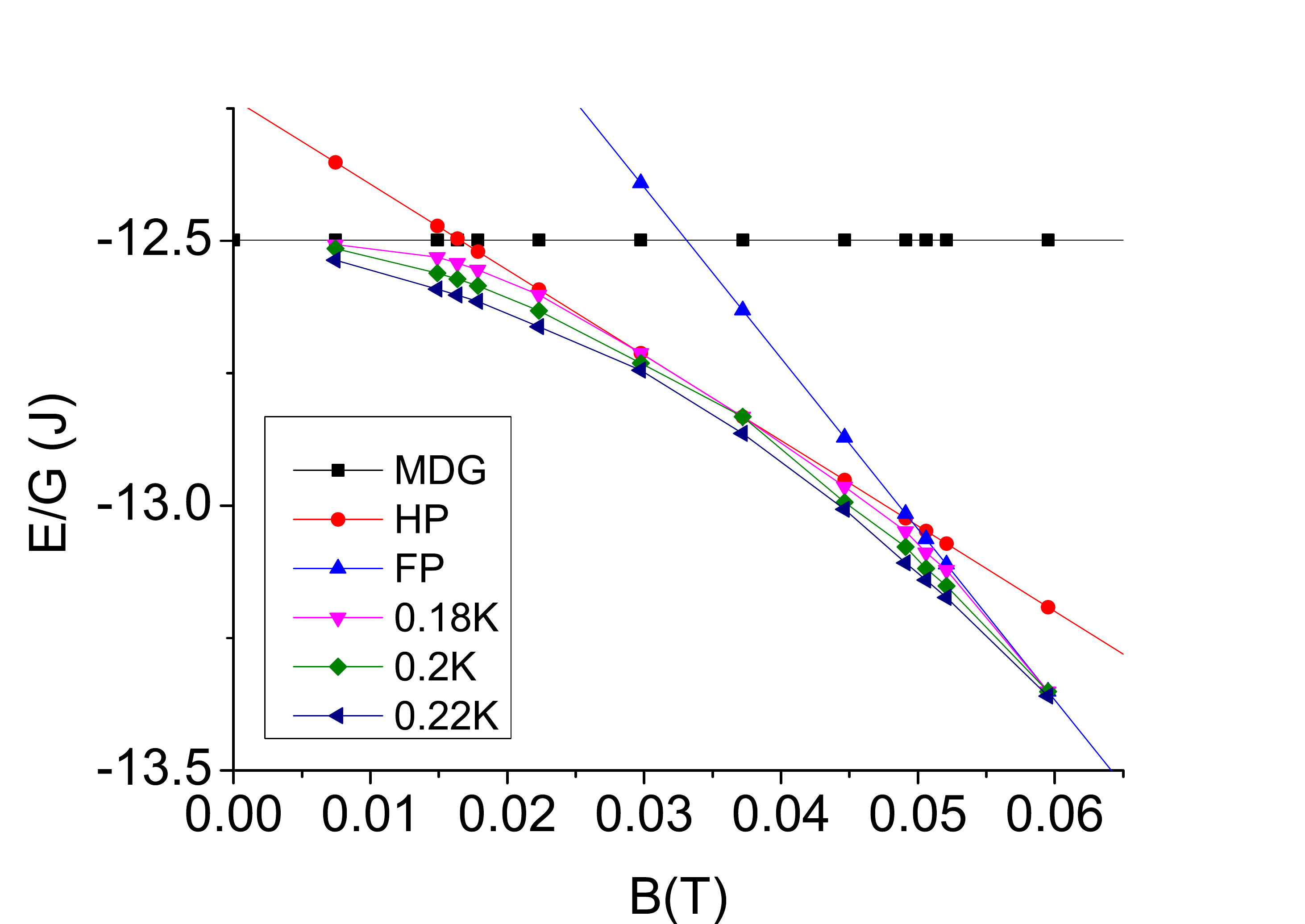}}
\caption{(Color online). Field dependence of the energy ($E$) per spin for the MDG phase (black square), the HP phase (red circle) and the FP (blue upper triangle). The level crossings correspond to the plateau transitions at zero temperature. Also plotted is the free energy ($G$) per spin at  $T$=0.22K (purple left triangle), $0.20$K (green diamond), and  $0.18$K (pink down triangle).}
\label{fig:free_energy}
\end{figure}

\section{Level crossings}
To better understand the temperature evolution of the field dependence of the magnetization, we examine the energy of the possible ordered states and the free energy density of the system at different temperatures  in Fig.~\ref{fig:free_energy}. First, we consider the field dependence of  the energy for the  ordered states:  the  MDG state (black square), the HP state (red circle), and the FP state (blue upper triangle). There exist two level crossings at $B=0.018$, and 0.052T respectively, which corresponds to the plateau transitions at low temperature. We also compute the the free energy $G=E-TS$ at  $T=0.18$K, 0.20K and 0.22K from the simulation data. The entropy is obtained by numerically integrating  the specific heat data. At $T=0.22$K (purple left triangle), none of the  ordered states has the lowest free energy and the system remains disordered.  At $0.20$K (green diamond),   the free energy curve touches the HS state  near $B=0.035$T, indicating the appearance of the half-magnetization plateau. At $0.18$K (pink down triangle), for $B<0.01$T, the MDG  state has the lowest free energy, for $B$=0.025 to  0.045T, the HP state is the ground state and for $B>0.052$T, the system enters the fully polarized state. Between these plateaus, the system remains in the disordered Coulomb phase. These results are consistent with what is observed in the field dependence of the magnetization (Fig.~3 in the main text).

\section{Specific heat and magnetic susceptibility}
\begin{figure}[tbp]
\centerline{\includegraphics[width=0.95\columnwidth,clip]{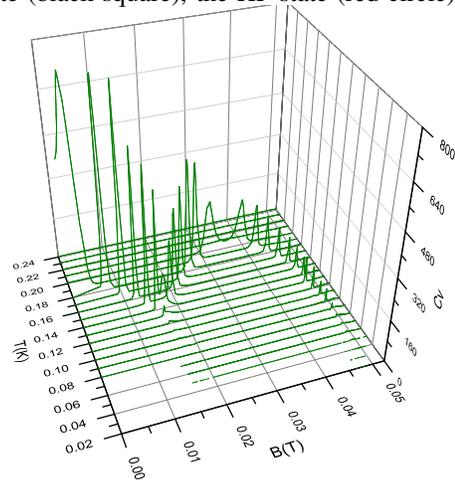}
}
\caption{(Color online.) The specific heat  data showing very sharp anomalies at the thermal transitions from the Coulomb to the ordered states. In contrast, relatively small  specific heat anomaly is observed for the low-temperature transitions between the ordered states.  }
\label{fig:cv}
\end{figure}

\begin{figure}[tbp]
\includegraphics[width=0.95\columnwidth,clip]{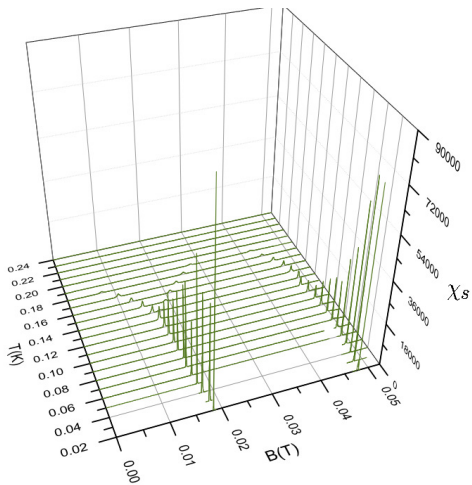}
\caption{ (Color online)
Magnetic susceptibility versus temperature and magnetic field. Strong magnetic susceptibility anomalies can be observed near the plateau transitions.
}
\label{fig:xs}
\end{figure}

Figure~\ref{fig:cv}  shows the specific heat data for $L = 5$. The thermal transitions from the spin ice state to the ordered states show  sharp peaks due to the release of the large residual entropy at the transition. In contrast, the transitions between the ordered states show relatively small specific heat anomaly as these are  ordered states with different magnetizations.
Figure~\ref{fig:xs} shows the magnetic susceptibility versus temperature and magnetic field for $L=5$. Strong magnetic susceptibility anomalies can be observed near the plateau transitions. This should be contrasted with the specific heat plot where it shows small signature of anomalies near the plateau transitions since the states are all ordered states with different magnetization. 

\bibliography{DipolarIce}

\end{document}